\documentclass[11pt,epsf,letterpaper]{article}%
\usepackage{color}
\usepackage{amsmath}
\usepackage{amsfonts}
\usepackage{verbatim}
\usepackage{amssymb}
\usepackage{graphicx}
\usepackage{epstopdf}
\usepackage{mathrsfs}%
\providecommand{\U}[1]{\protect\rule{.1in}{.1in}}
\DeclareMathOperator{\Tr}{Tr}
\begin{document}

\title{\bf{Study of Yang--Mills--Chern--Simons theory in presence of the Gribov horizon}}
\author{$^{(1,2)}$Fabrizio Canfora\thanks{canfora@cecs.cl}, $^{(3)}$Arturo Gomez\thanks{arturo.gomez@proyectos.uai.cl}\, $^{(4)}$Silvio Paolo Sorella\thanks{Work supported by
FAPERJ, Funda{\c{c}}{\~{a}}o de Amparo {\`{a}} Pesquisa do Estado do Rio de
Janeiro, under the program \textit{Cientista do Nosso Estado }, E-26/101.578/2010. sorella@uerj.br},\; $^{(4)}$David Vercauteren\thanks{vercauteren.uerj@gmail.com}
\\\\\textit{$^{(1)}$ Centro de Estudios Cientificos (CECs), Valdivia,  Chile}\\\\\textit{$^{(2)}$ Universidad Andres Bello, Av. Republica 440, Santiago, Chile}\\\\\textit{$^{(3)}$Departamento de Ciencias, Facultad de Artes Liberales y}\\\textit{Facultad de Ingenier\'{\i}a y Ciencias, Universidad Adolfo
Ib\'{a}\~{n}ez, Vi\~{n}a del Mar.}\\\\\textit{$^{(4)}$UERJ, Universidade do Estado do Rio de Janeiro (UERJ),} \\\textit{ Instituto de F\'{\i}sica Te\'orica, Rua S\~ao Francisco Xavier 524,}\\\textit{20550-013, Maracan\'a, Rio de Janeiro, Brasil }}
\maketitle

\begin{abstract}
The two-point gauge correlation function in Yang--Mills--Chern--Simons theory in three dimensional Euclidean space is analysed by taking into account the non-perturbative effects of the Gribov horizon. In this way, we are able to describe the  confinement and de-confinement regimes, which naturally depend on the topological mass and on the gauge coupling constant of the theory.  

\end{abstract}

\section{Introduction}

Three-dimensional Yang--Mills theory is one of the most important models in
which it is possible to analyse unsolved non-perturbative problems such as
color confinement. The theory is  simpler than QCD, but it is still highly
non-trivial. It has local degrees of freedom and the coupling constant is dimensionful. 
Moreover, it can be viewed as an approximation to the high temperature phase
of QCD with the mass gap serving as the magnetic mass. \\\\A very interesting term which can be added to the three-dimensional Yang--Mills theory is the
Chern--Simons term \cite{Deser:1982vy,Deser:1981wh}\footnote{The Chern--Simons term can arise, for instance,
from the evaluation of the fermionic determinant in three dimensions (see, for a detailed review, \cite{dunne}). See also  \cite{Yildirim:2013cna} for a recent application of Yang-Mills-Chern-Simons theory to the theory of the link invariants.}: this term provides a mass for the gauge field which is of topological origin. Therefore, while pure $3d$ Yang--Mills is known to be a confining theory, the addition of the topological Chern--Simons term has the effect of generating a de-confined massive excitation. Said otherwise, the theory undergoes a change of regime, passing from a confined to a de-confined regime. \\\\The purpose of this paper is that of discussing, within a quantum field theory framework, how this change of regime is driven by the presence of the Chern--Simons term. To that aim, we shall take into account the non-perturbative effects arising from the Gribov horizon \cite{Gribov:1977wm}\footnote{For a pedagogical introduction to the Gribov problem, see refs.\cite{Sobreiro:2005ec,Vandersickel:2012tz}.}. This will enable us to encode non perturbative effects into the two-point gluon correlation function whose analytic structure can be employed to analyse how the theory moves from one regime to another when varying the coupling constant $g$ and the Chern--Simons mass parameter $M$. We remind here that the presence of the Gribov phenomenon is a general feature of the quantization procedure of  nonabelian gauge theories, the existence of Gribov copies being in fact a well known property of any local covariant renormalizable gauge fixing  \cite{Singer:1978dk} (see also \cite{jackiwmuz}). The presence of gauge copies gives rise to zero modes of the Faddeev--Popov operator which  invalidate the usual Faddeev--Popov construction. \\\\A successful method to deal with the issue of the Gribov copies is that of restricting the domain of integration in the functional integral to the so-called Gribov region $\Omega$ \cite{Gribov:1977wm,Sobreiro:2005ec,Vandersickel:2012tz}, which is the set of all transverse field configurations for which the Faddeev--Popov operator ${\cal M}^{ab}=-\partial_\mu D_{\mu}^{ab}$ is strictly positive, namely $\Omega
=\{A_{\mu}^{a};\;\;\partial_{\mu}A_{\mu}^{a}=0,\;\;{\cal M}^{ab}>0\}$. The region $\Omega$ has been proven to be bounded in all directions in field space \cite{Dell'Antonio:1989jn}, its boundary $\partial \Omega$ being the first Gribov horizon. Moreover, all gauge orbits pass through $\Omega$ at least once \cite{Dell'Antonio:1991xt}, a property which strongly supports the restriction to $\Omega$. Remarkably, the whole procedure results in a local and renormalizable action known as the Gribov--Zwanziger action \cite{Zwanziger:1989mf,Zwanziger:1992qr}. More recently, a refinement of the Gribov--Zwanziger action has been worked out in  \cite{Dudal:2007cw,Dudal:2008sp} by taking account the effects of dimension two condensates. The resulting two-point gluon correlation function turns out to be in excellent agreement with the most recent lattice data \cite{Cucchieri:2011ig}, allowing for nontrivial analytic estimates of the first glueball states \cite{Dudal:2010cd,Dudal:2013wja}. Let us also mention that the Refined Gribov--Zwanziger framework has been employed in the study of the Casimir energy \cite{Canfora:2013zna}, producing the correct sign for the Casimir force within the MIT bag model, clarifying a long-standing problem. Also, in a series of papers   \cite{Capri:2012cr,Capri:2012ah,Capri:2013gha,Capri:2013oja}, the Gribov--Zwanziger set up has been employed in order to study, in the continuum, the transition between the confining and non-confining regimes when Higgs fields are present. Also in this case, the non-perturbative gluon two-point correlation function  obtained by taking into account the effects of the Gribov horizon turns out to be a useful quantity in order to obtain information about the transition from the confining to the Higgs regime. As discussed in details in   \cite{Capri:2012cr,Capri:2012ah,Capri:2013gha,Capri:2013oja}, the gluon correlation function undergoes a continuous change from a confining expression of the Gribov type, characterized by the presence of unphysical complex conjugate poles, to a Yukawa type propagator with a real pole, indicating that the theory is in the Higgs regime.  The emerging picture is in full agreement with the renewed Fradkin--Shenker work  \cite{Fradkin:1978dv}.  \\\\In the present paper,  we shall implement the restriction to the Gribov region $\Omega$  in $3d$ Yang--Mills--Chern--Simons theory by working out the non-perturbative expression of the two point gauge correlation function. Further, we shall vary the gauge coupling constant $g$ and the Chern--Simons mass $M$ and discuss how the poles of this correlation function get modified, thus obtaining information on how the theory passes form the confining to the non-confining regimes. \\\\The paper is organized as follows: in Section 2, the gluon propagator and the Gribov gap equation for $3d$ Yang--Mills--Chern--Simons  theory are obtained. In Section 3, the behaviour of the poles
of the gauge  propagator as functions of the two parameters $(g,M) $ is discussed. In Section IV we present our conclusions.  

\section{Gauge propagator for  Yang--Mills--Chern--Simons action in presence of the Gribov horizon}
We start by considering the Yang--Mills--Chern--Simons action in $3d$ Euclidean flat space quantized in the Landau gauge, namely 
\begin{equation}
S_{M}=-iM\int d^{3}x\;\epsilon_{\mu\rho\nu}\left(  \frac{1}{2}A_{\mu}%
^{a}\partial_{\rho}A_{\nu}^{a}+\frac{1}{3!}gf^{abc}A_{\mu}^{a}A_{\rho}%
^{b}A_{\nu}^{c}\right) +\frac{1}{4}\int d^{3}x\;F_{\mu\nu}^{a}F_{\mu\nu}%
^{a}+\int d^{3}x\left(  b^{a}\partial_{\mu}A_{\mu}^{a}+\bar{c}^{a}%
\partial_{\mu}D_{\mu}^{ab}c^{b}\right)   \label{SYMCS}%
\end{equation}
Here, $M$ stands for the Chern--Simons mass, $b^a$ is the Lagrange multiplier enforcing the Landau gauge, $\partial_\mu A^a_\mu=0$, and $({\bar c}^a, c^a)$ are the Faddeev--Popov ghosts.  This theory is known as the topologically massive non-Abelian gauge theory,
because of the massive gluon propagator \cite{Deser:1982vy,Deser:1981wh}, given by 
\begin{equation}
\left\langle A_{\mu}^{a}(q)A_{\mu}^{a}(-q)\right\rangle =\frac{\delta^{ab}%
}{(q^{2}+M^{2})}\left(  \delta_{\mu\nu}-\frac{q_{\mu}q_{\nu}}{q^{2}}%
+M\epsilon_{\mu\rho\nu}\frac{q_{\rho}}{q^{2}}\right)  \ , \label{PropCS}%
\end{equation}
As already mentioned in the Introduction, the action \eqref{SYMCS} is plagued by the existence of Gribov copies. We shall thus proceed by restricting the domain of integration in the functional integral to the Gribov region $\Omega$. To that aim we shall follow the procedure outlined by Gribov \cite{Gribov:1977wm,Sobreiro:2005ec,Vandersickel:2012tz}.  It amounts to impose the so-called  no-pole condition for the connected two-point ghost function $\mathcal G^{ab} (k;A) = \langle k | (-\partial D^{ab}(A))^{-1} | k \rangle$, which is nothing but the inverse of the Faddeev--Popov operator $-\partial_\mu D_\mu^{ab}(A)$. One requires that $\mathcal G^{ab}(k;A)$ has no poles at finite non-vanishing values of $k^2$, so that it stays always positive. In that way one ensures that the Gribov horizon is not crossed, {\it i.e.} one remains inside $\Omega$. The only allowed pole is at $k^2 = 0$, which has the meaning of approaching the boundary of the region $\Omega$. \\\\Following Gribov's procedure \cite{Gribov:1977wm,Sobreiro:2005ec,Vandersickel:2012tz}, for the connected two-point ghost function $\mathcal G^{ab}(k;A)$ at first
order in the gauge fields, one finds
\begin{equation}
\mathcal{G}^{ab}(k;A) = \frac{1}{k^{2}}\left( \delta^{ab} - g^{2}\frac{k_{\mu}k_{\nu}}{k^{2}} \int \frac{d^{3}p}{(2\pi)^{3}} f^{ace} f^{dbe} \frac{A^c_{\mu}(p)A^d_{\nu}(-p)}{(k-p)^{2}}\right) \; .
\label{ghprop}
\end{equation}
One can then write
\begin{equation}
\mathcal G(k;A) = \frac{1}{N^2-1} \sum_{a=1}^{N^2-1}\mathcal G^{aa}(k;A) \approx \frac1{k^2} \frac1{1-\sigma(k;A)} \;, \label{defG}
\end{equation}
where the form factor $\sigma(k;A)$ is given by 
\begin{equation}
	\sigma(k;A) = \frac{Ng^{2}}{N^2-1} \frac{k_{\mu}k_{\nu}}{k^{2}} \int \frac{d^{3}p}{(2\pi)^{3}} \frac{A^a_{\mu}(p)A^a_{\nu}(-p)}{(k-p)^{2}} \;. \label{ffactor}
\end{equation}
The quantity $\sigma(k;A)$ turns out to be a decreasing function of the momentum $k$ \cite{Gribov:1977wm,Sobreiro:2005ec,Vandersickel:2012tz}. Thus, the no-pole condition is implemented by requiring that \cite{Gribov:1977wm,Sobreiro:2005ec,Vandersickel:2012tz}
\begin{equation}
	\sigma(0;A) \leq 1 \;.   \label{nopole}
\end{equation}
Making use of the transversality of $A^a_{\mu}(p)A^a_{\nu}(-p)$ in the Landau gauge, one easily  finds that
\begin{equation}
	\sigma(0;A) = \frac{Ng^{2}}{N^2-1} \frac1d \int \frac{d^{3}p}{(2\pi)^{3}} \frac{A^a_{\mu}(p)A^a_{\mu}(-p)}{p^{2}}
\end{equation}
One can easily verify that this condition is already fulfilled \emph{without} restriction to the Gribov horizon whenever
\begin{equation} \label{nogribov}
	\frac{Ng^2}{6\pi M} < 1 \;.
\end{equation}
This means that, in the weak coupling regime, the Gribov problem does not occur. \\\\ Although in the present paper we are mainly focusing on the Gribov copies related to infinitesimal gauge transformations, namely to copies related to zero modes of the Faddeev-Popov operator, it is worth remanding here that, as pointed out in  \cite{Deser:1982vy,Deser:1981wh}, the Chern-Simons term is not left invariant by the so called large gauge transformations, 
{\it i.e.} gauge transformations with non-vanishing winding number. Nevertheless, gauge invariance of the partition 
function is achieved by demanding that the Chern-Simons mass $M$ obeys a quantization rule. More precisely, from  
 \cite{Deser:1982vy,Deser:1981wh} one has that $4\pi \frac{M}{g^2} = n$, where $n$ is an integer, $n \pm 1, \pm 2, \cdots$. Therefore, combining this
quantization rule with expression (8), one learns that for values of the integer $n$ such that $n> \frac{2}{3} N$, the size 
of the Chern-Simons mass $M$ still guarantees that the no-pole condition (6) is fulfilled.  To some extent, this remark might give a first indication of whether the Gribov copies related to large gauge transformations are expected to be not relevant.  \\\\As done in \cite{Gribov:1977wm,Sobreiro:2005ec,Vandersickel:2012tz}, condition \eqref{nopole} is encoded in the Euclidean functional measure through the introduction of a step function $\theta(x)$. Therefore, for the partition function of the theory one gets
\begin{eqnarray}
Z & = & \int [DA]\; \delta(\partial A^a) \; \det(-\partial D^{ab}) \; \theta(1-\sigma(0;A)) \; e^{-\left(S_{YM} + S_{CS} \right)}  \; , \label{Zc} \\
S_{YM}  & = & \frac{1}{4}\int d^{3}x\;F_{\mu\nu}^{a}F_{\mu\nu}^{a}  \; , \nonumber \\ 
S_{CS} & = & -iM\int d^{3}x\;\epsilon_{\mu\rho\nu}\left(  \frac{1}{2}A_{\mu}%
^{a}\partial_{\rho}A_{\nu}^{a}+\frac{1}{3!}gf^{abc}A_{\mu}^{a}A_{\rho}%
^{b}A_{\nu}^{c}\right)  \nonumber \;.
\end{eqnarray}
The factor $\sigma(0;A)$ can be lifted into the exponential by employing the following integral representation for the step function 
\begin{equation}
	\theta(1-\sigma(0;A)) = \int_{-i\infty+\epsilon}^{+i\infty+\epsilon} \frac{d\beta}{2\pi i\beta} e^{\beta(1-\sigma(0;A))} \;, 
\end{equation}
so that 
\begin{equation}
Z  =  \int [DA] \; \frac{d\beta}{2\pi i\beta}\; \delta(\partial A^a) \; \det(-\partial D^{ab})  \; e^{-\left(S_{YM} + S_{CS} \right)-\beta(1-\sigma(0;A))}   \label{Zca} \;.
\end{equation}
As, in the following, we are concerned with the gauge propagator, we shall focus on the partition function in the quadratic approximation, namely
\begin{equation}
Z_{quad}=\int \frac{d\beta }{2\pi i\beta } \; [DA] \; e^{\beta}\;e^{-\frac{1}{2}\int \frac{d^3q}{(2\pi)^3} A_{\mu
}^{a}(q) Q_{\mu \nu }^{ab}A_{\nu }^{b}(-q)} \ ,  \label{Zq}
\end{equation}
with
\begin{equation}
Q_{\mu\nu}^{ab}    =\delta^{ab}\left(  q^{2}\delta_{\mu\nu}+\left(  \frac
{1}{\alpha}-1\right)  q_{\mu}q_{\nu}+\frac{\gamma^{4}}{q^{2}}\delta_{\mu\nu
}-M\epsilon_{\mu\rho\nu}q_{\rho}\right)  \;, \label{quad}
\end{equation}
where $\alpha$ is a gauge parameter to be set to zero after having evaluated the gauge propagator. The parameter $\gamma$ stands for the Gribov mass parameter \cite{Gribov:1977wm,Sobreiro:2005ec,Vandersickel:2012tz}:
\begin{equation} 
\gamma^{4}    =\frac{2}{3} \frac{N g^2}{N^{2}-1}\beta \;.  \label{gamma}
\end{equation}
In order to evaluate the gauge propagator it suffices to invert the operator  $Q_{\mu\nu}^{ab}$. Writing
\begin{equation}
\left(  Q_{\mu\nu}^{ab}\right)  ^{-1}=\delta^{ab}\left(  F(q)\delta_{\mu\nu
}+B(q)q_{\mu}q_{\nu}+C(q)\epsilon_{\mu\nu\rho}q_{\rho}\right)  \ ,
\label{inverse}%
\end{equation}
the coefficients $F,B$ and $C$ are determined by requiring that
\begin{equation}
Q_{\mu\nu}^{ab}\left(  Q_{\nu\lambda}^{bc}\right)  ^{-1}=\delta^{ac}%
\delta_{\mu\lambda}\;, \label{identity}%
\end{equation}
yielding the following expression for the gauge propagator
\begin{equation}
\left\langle A_{\mu}^{a}(q)A_{\nu}^{b}(-q)\right\rangle =\delta^{ab}%
\frac{q^{2}\left(  q^{4}+\gamma^{4}\right)  }{\left(  q^{4}+\gamma^{4}\right)
^{2}+M^{2}q^{6}}\left(  \delta_{\mu\nu}-\frac{q_{\mu}q_{\nu}}{q^{2}}%
+\frac{q^{2}}{q^{4}+\gamma^{4}}M\epsilon_{\mu\lambda\nu}q_{\lambda}\right)
\label{YMCS-GribovP} \;.
\end{equation}
It is worth noting that, removing the Gribov horizon, {\it i.e.} setting   $\gamma=0$ in eq. (\ref{YMCS-GribovP}), we recover
the massive propagator of Yang--Mills--Chern--Simons theory, eq. \eqref{PropCS}.  On the other hand,
when $M=0$, the Gribov propagator for Yang--Mills theory is obtained, that is 
\begin{equation}
\left\langle A_{\mu}^{a}(q)A_{\nu}^{b}(-q)\right\rangle_{ Gribov}  =\delta^{ab}
\frac{q^{2}}{\left(  q^{4}+\gamma^{4}\right) }\left(  \delta_{\mu\nu}-\frac{q_{\mu}q_{\nu}}{q^{2}} \right)  \;.
\label{GribovP}
\end{equation}

\subsection{The gap equation for the Gribov parameter $\gamma$}

The Gribov parameter $\gamma$ is not free, being  determined in a self-consistent way through a suitable 
gap equation, which we shall derive below by following Gribov's setup, amounting to evaluate the partition function \eqref{Zq}  at the saddle point \cite{Gribov:1977wm,Sobreiro:2005ec,Vandersickel:2012tz}. To that end we write 
$Z_{quad}$ as 
\begin{equation}
Z_{quad} = \int \frac{d\beta }{2\pi i\beta } \; e^{-f(\beta) }\;, \label{zq1}
\end{equation}
where, after integrating out the gauge fields, the quantity $f(\beta)$ is given by 
\begin{equation}
f(\beta) = \frac{1}{2} \Tr \ln Q_{\mu\nu}^{ab} + \ln\beta -\beta  \;. \label{fbeta}
\end{equation}
In the Gribov semiclassical approximation \cite{Gribov:1977wm,Sobreiro:2005ec,Vandersickel:2012tz}, expression \eqref{zq1} is evaluated at the 
saddle point, {\it i.e.}
\begin{equation}
Z_{quad} \simeq \; e^{-f(\beta^*)}  \;, \label{zq2} 
\end{equation}
where $\beta^*$ corresponds to the stationary point of $f(\beta)$
\begin{equation} 
 \frac{\partial f(\beta)} {\partial \beta} \Bigl|_{\beta=\beta^*}
= 0 \;, \label{stat}
\end{equation} 
which, upon evaluating $\Tr \ln Q_{\mu\nu}^{ab}$, gives the gap equation for the Gribov parameter 
$\gamma$:
\begin{equation}
\frac{2Ng^{2}}3 \int\frac{d^{3}q}{(2\pi)^{3}}\frac{1}{q^{4}+\gamma^{4}%
}=1 \;.\label{Gap}%
\end{equation}
It is worth noticing that the Chern--Simons mass parameter $M$ does not enter the gap equation \eqref{Gap}. This is an expected result, due to the topological nature of the Chern--Simons term. One recognizes in fact  that the quantity $f(\beta^*)$ has the physical meaning of the vacuum energy of the system. However, as the Chern--Simons term does not couple to the metric, it follows that it does not contribute to the vacuum energy, which turns out to be independent from $M$. Of course, the same happens with the gap equation \eqref{Gap} for the parameter $\gamma$. Nevertheless, the presence of the Chern--Simons term leads to a deep change of
the structure of the gauge propagator.  \\\\It is straightforward  to integrate equation $\left(  \ref{Gap}\right)  $, obtaining $\gamma$ as a function of the coupling constant $g$, {\it i.e.}  
\begin{equation}
\gamma=\lambda^{1/4} g^{2}\;, \qquad \lambda^{1/4}=\frac{\sqrt2N}{12\pi} \;. %
\end{equation}
Therefore, the gauge propagator takes the  form
\begin{equation}
\left\langle A_{\mu}^{a}(q)A_{\nu}^{b}(-q)\right\rangle =\delta^{ab}%
\frac{q^{2}\left(  q^{4}+\lambda g^{8}\right)  }{\left(  q^{4}+\lambda
g^{8}\right)  ^{2}+M^{2}q^{6}}\left(  \delta_{\mu\nu}-\frac{q_{\mu}q_{\nu}%
}{q^{2}}+\frac{q^{2}}{q^{4}+\lambda g^{8}}M\epsilon_{\mu\lambda\nu}q_{\lambda
}\right)  \label{propfinal1}%
\end{equation}
We notice that, before the implementation of the restriction to the Gribov region $\Omega$, the theory displays a massive Yukawa type mode, as it is apparent
from the expression of the propagator in eq.$\left(  \ref{PropCS}\right)$.  The question that
naturally arises is under which conditions this physical mode survives when the influence of
the Gribov horizon is taken into account. This is the topic we shall address in the next section.

\section{Analytic structure of the gauge propagator and the different regimes of the theory}
The propagator in expression $\left(  \ref{propfinal1}\right)  $ depends on
the coupling constant $g$ and on the Chern Simons mass
$M$, and exhibits a rather complex pole structure. The poles of the propagator are functions of the parameters $(g,M)$. As such, the study of their behavior when varying $(g,M)$ is of great help in understanding the different regimes in which the theory may be found, as recently discussed in the case of Yang--Mills theories in the presence of Higgs fields \cite{Capri:2012cr,Capri:2012ah,Capri:2013gha,Capri:2013oja}  as well as of gauge theories at finite temperature \cite{Canfora:2013kma}. The region in the plane $(g,M)$ in which the poles of the gauge propagator are complex has a natural interpretation as a confining region, since complex poles cannot be associated to a physical excitation of the spectrum. Moreover, the region in which the poles are real and the corresponding residues are positive has the meaning of a deconfined region in which a massive gauge particle is present in the spectrum. \\\\In order to find the poles of the propagator \eqref{propfinal1} we have to determine the roots of the following polynomial
\begin{subequations} \begin{align}
P(q^2) &  =\left(  q^{4}+\lambda g^{8}\right)^{2}+M^{2}q^{6}  \label{P(x)} \\
&  =q^{8}+M^{2}q^{6}+2Gq^{4}+G^{2}   \text{ \ with }G=\lambda g^{8}\nonumber\\
&  =(q^2+m_{1}^2)(q^2+m_{2}^2)(q^2+m_{3}^2)(q^2+m^2_{4}) \;, \label{decomposition}%
\end{align} \end{subequations}
where $G=\lambda g^{8}$. Although in principle the roots $(m^2_1, m^2_2,m^2_3,m^2_4)$ can be evaluated in close form, they turn out to be complicated expression of the parameters $(G,M)$. Rather, in order to provide  a better and more clear analysis, we shall display three dimensional plots of the poles as functions of $(G,M)$. Let us start by splitting the 
propagator $\left(  \ref{propfinal1}%
\right)  $ in two parts, a parity conserved, and a parity violating one, namely 
\begin{equation}
\left\langle A_{\mu}^{a}(q)A_{\nu}^{b}(-q)\right\rangle =\left.
\mathcal{G}_{\mu\nu}^{ab}\mathcal{(}q\mathcal{)}\right\vert _{par}+\left.
\mathcal{G}_{\mu\nu}^{ab}\left(  q\right)  \right\vert _{par-viol}
\end{equation} \label{decomposition1}
with
\begin{align}
\left.  \mathcal{G}_{\mu\nu}^{ab}\mathcal{(}q\mathcal{)}\right\vert _{par}  &
=\delta^{ab}\frac{q^{2}\left(  q^{4}+\lambda g^{8}\right)  }{\left(
q^{4}+\lambda g^{8}\right)  ^{2}+M^{2}q^{6}}\left(  \delta_{\mu\nu}%
-\frac{q_{\mu}q_{\nu}}{q^{2}}\right) \;,  \label{Gpar}\\
\left.  \mathcal{G}_{\mu\nu}^{ab}\left(  q\right)  \right\vert _{par-viol}  &
=\delta^{ab}\frac{q^{4}}{\left(  q^{4}+\lambda g^{8}\right)  ^{2}+M^{2}q^{6}%
}M\epsilon_{\mu\lambda\nu}q_{\lambda} \label{Gimpar} \;.
\end{align}
Using partial fraction decomposition,  for the parity violating part of the gluon propagator  we get 
\begin{equation}
\left.  \mathcal{G}_{\mu\nu}^{ab}\left(  q\right)  \right\vert _{non-par}%
=\delta^{ab}\left(  \frac{\mathcal{R}_{1}}{q^{2}+m_{1}^{2}}+\frac
{\mathcal{R}_{2}}{q^{2}+m_{2}^{2}}+\frac{\mathcal{R}_{3}}{q^{2}+m_{3}^{2}%
}+\frac{\mathcal{R}_{4}}{q^{2}+m_{4}^{2}}\right)  M\epsilon_{\mu\lambda\nu
}q_{\lambda} \label{nonparpoles}%
\end{equation}
where the residues $\mathcal{R}_{1},\mathcal{R}_{2},\mathcal{R}_{3},\mathcal{R}_{4}$ are given by 
\begin{align}
\mathcal{R}_{1}  &  =\frac{m_{1}^{4}}{(m_{2}^{2}-m_{1}^{2})(m_{3}^{2}%
-m_{1}^{2})(m_{4}^{2}-m_{1}^{2})}\label{R1} \;, \\
\mathcal{R}_{2}  &  =-\frac{m_{2}^{4}}{(m_{2}^{2}-m_{1}^{2})(m_{3}^{2}%
-m_{2}^{2})(m_{4}^{2}-m_{2}^{2})}\label{R2} \;, \\
\mathcal{R}_{3}  &  =\frac{m_{3}^{4}}{(m_{1}^{2}-m_{3}^{2})(m_{2}^{2}%
-m_{3}^{2})(m_{4}^{2}-m_{3}^{2})}\label{R3}\;, \\
\mathcal{R}_{4}  &  =-\frac{m_{4}^{4}}{(m_{4}^{2}-m_{1}^{2})(m_{4}^{2}%
-m_{3}^{2})(m_{4}^{2}-m_{2}^{2})}\label{R4}%
\end{align}
Analogously,  the parity conserved part $\left(
\ref{Gpar}\right)  \ $ of the gluon propagator reads

\begin{equation}
\left.  \mathcal{G}_{\mu\nu}^{ab}\mathcal{(}q\mathcal{)}\right\vert
_{par}=\delta^{ab}\left(  \frac{{\cal{F}}_{1}}{q^{2}+m_{1}^{2}}%
+\frac{{\cal{F}}_{2}}{q^{2}+m_{2}^{2}}+\frac{{\cal{F}}_{3}}{q^{2}+m_{3}^{2}%
}+\frac{{\cal{F}}_{4}}{q^{2}+m_{4}^{2}}\right)  \left(  \delta_{\mu\nu}%
-\frac{q_{\mu}q_{\nu}}{q^{2}}\right)  \label{parpoles}%
\end{equation}
and the residues are,%
\begin{align}
{\cal{F}}_{1} &  =\frac{m_{1}^{2}\left(  \mathit{G}+m_{1}^{4}\right)
}{(m_{2}^{2}-m_{1}^{2})(m_{3}^{2}-m_{1}^{2})(m_{4}^{2}-m_{1}^{2})}\label{r1} \;, \\
{\cal{F}}_{2} &  =-\frac{m_{2}^{2}\left(  \mathit{G}+m_{2}^{4}\right)
}{(m_{2}^{2}-m_{1}^{2})(m_{3}^{2}-m_{2}^{2})(m_{4}^{2}-m_{2}^{2})}\label{r2} \;, \\
{\cal{F}}_{3} &  =\frac{m_{3}^{4}\left(  \mathit{G}+m_{3}^{4}\right)
}{(m_{1}^{2}-m_{3}^{2})(m_{2}^{2}-m_{3}^{2})(m_{4}^{2}-m_{3}^{2})}\label{r3}\;, \\
{\cal{F}}_{4} &  =-\frac{m_{4}^{2}\left(  \mathit{G}+m_{4}^{4}\right)
}{(m_{1}^{2}-m_{4}^{2})(m_{3}^{2}-m_{4}^{2})(m_{4}^{2}-m_{2}^{2})}\;.\label{r4}%
\end{align}
 Let us discuss these results in more detail. Due to the dependence of the polynomial $P(q^2)$ in eq.\eqref{decomposition} on the parameters $M$ and $G$, the reality of the roots will depend on these parameters as well, possibly becoming complex, and thus turning the related mode unphysical, for certain values of them. Computing the discriminant of the quartic polynomial in \eqref{decomposition}
\begin{equation}
	\Delta = 256 M^4G^5 - 27 M^8G^4
\end{equation}
it is possible to discern a regime with four complex masses\footnote{That the four roots are complex and not real is easily found by explicitly computing them.} ($\Delta > 0$ or $G > (3M/4)^4$) from a regime with two complex and two real masses ($\Delta < 0$ or $G < (3M/4)^4$). This result can be interpreted as follows: At small values of the Chern--Simons mass $M$ and for large values of the coupling constant $G$ all excitation in the theory are confined. For large values of the Chern--Simons mass and weak coupling $G$ real poles appear in the propagator and the theory becomes deconfined. In this deconfined regime we can, furthermore distinguish two parts: one with intermediate values for the parameters ($M^4/4 < G < (3M/4)^4$) where there are two massive poles, and a weak-coupling regime with large Chern--Simons mass ($G < M^4/4$, see condition \eqref{nogribov}) where the Gribov problem does not occur and where the theory reduces to the perturbative behavior. 

\begin{figure}
\centering
\includegraphics{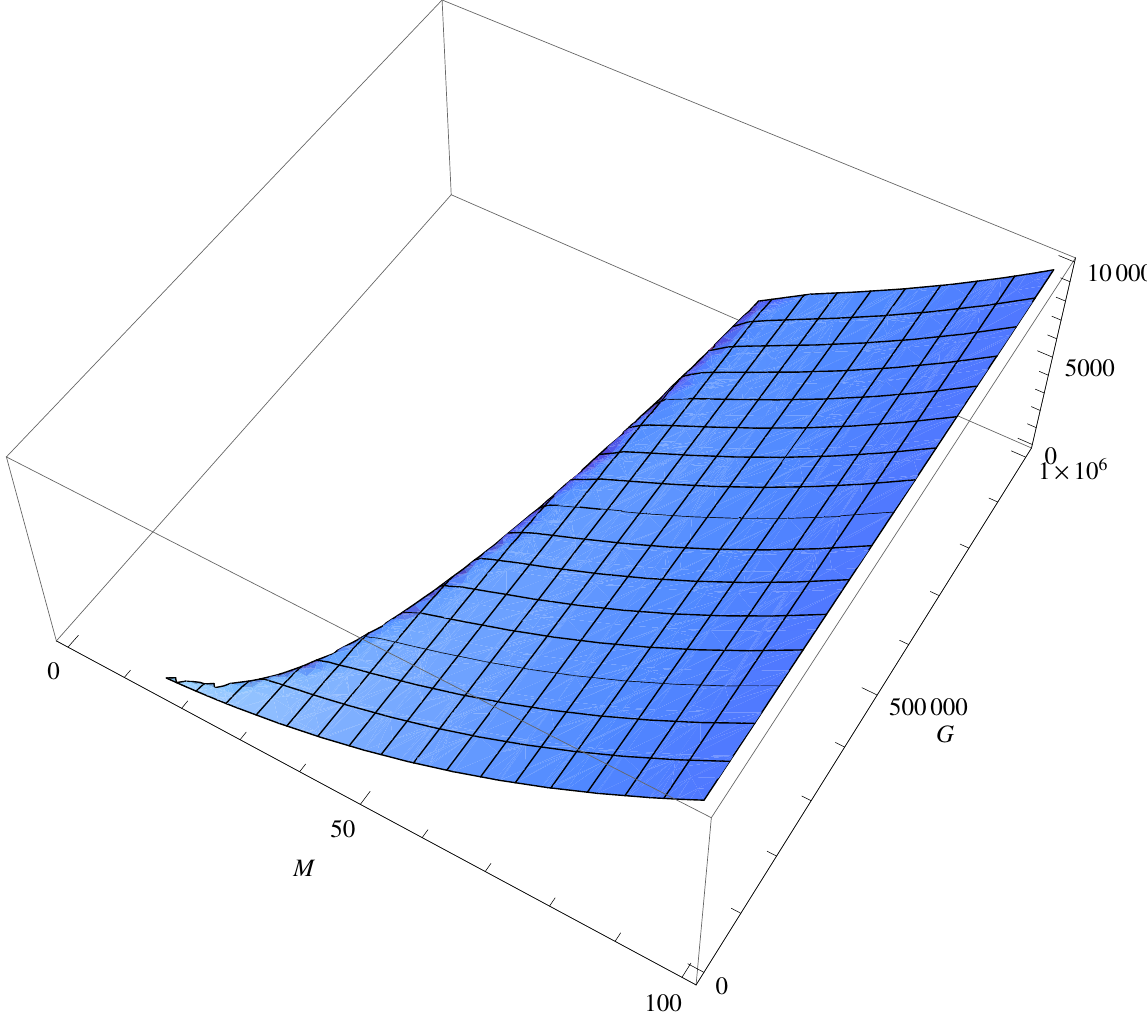}
\caption{$m_1$ as a function of the CS mass $M$ and $G=\lambda g^8$ \label{mone}}
\label{mass1}
\end {figure}

\begin{figure}
\centering
\includegraphics{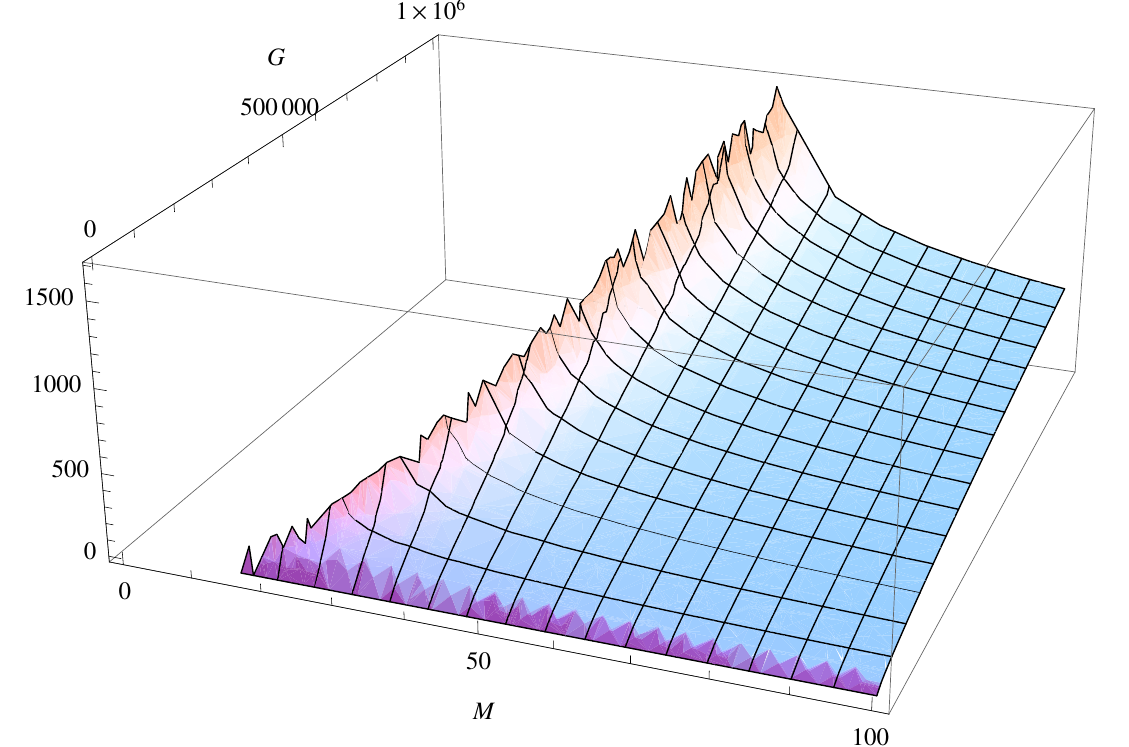}
\caption{$m_2$ as a function of the CS mass $M$ and $G=\lambda g^8$ \label{mtwo}}
\label{mass2}
\end {figure}

\begin{figure}
\centering
\includegraphics{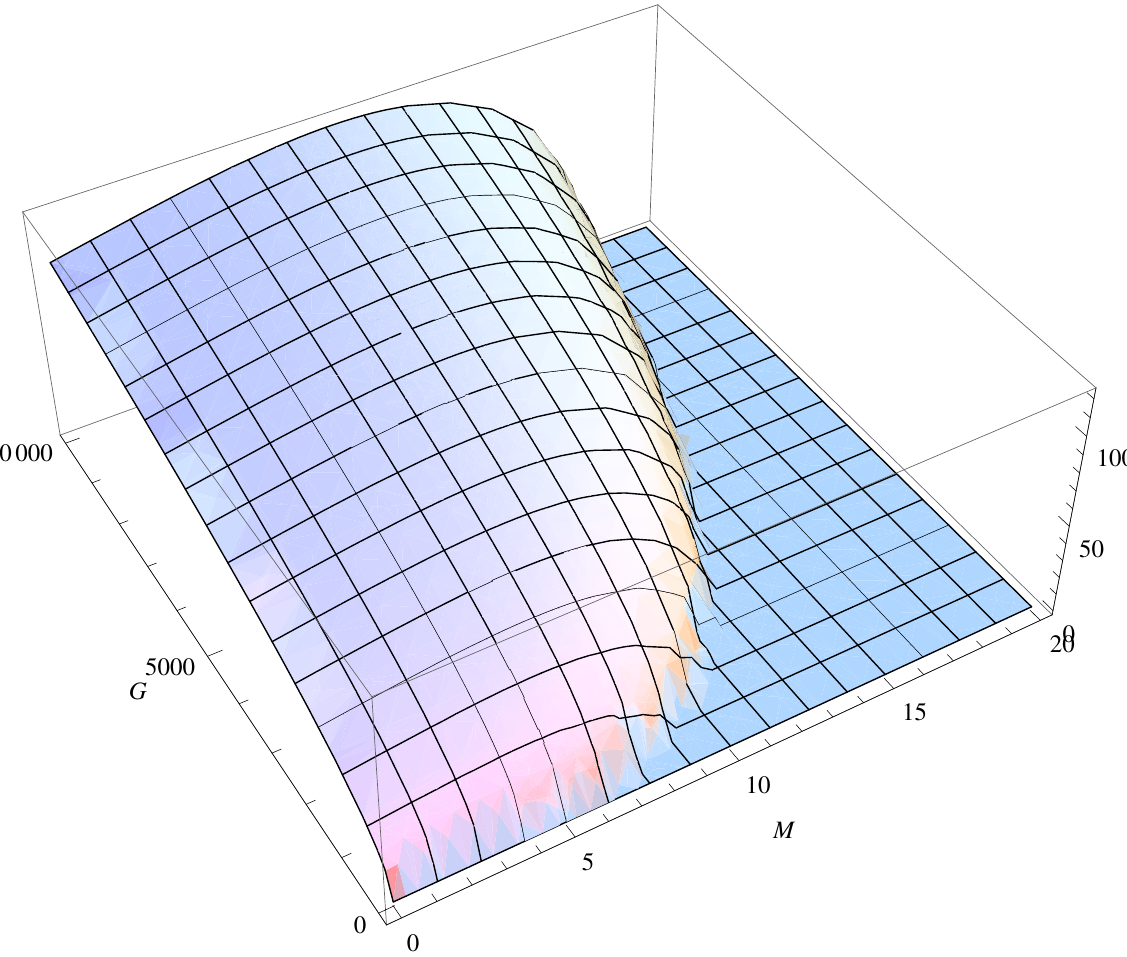}
\caption{Imaginary part of $m_1$ as a function of the CS mass $M$ and $G=\lambda g^8$ \label{imone}}
\end {figure}

\begin{figure}
\centering
\includegraphics{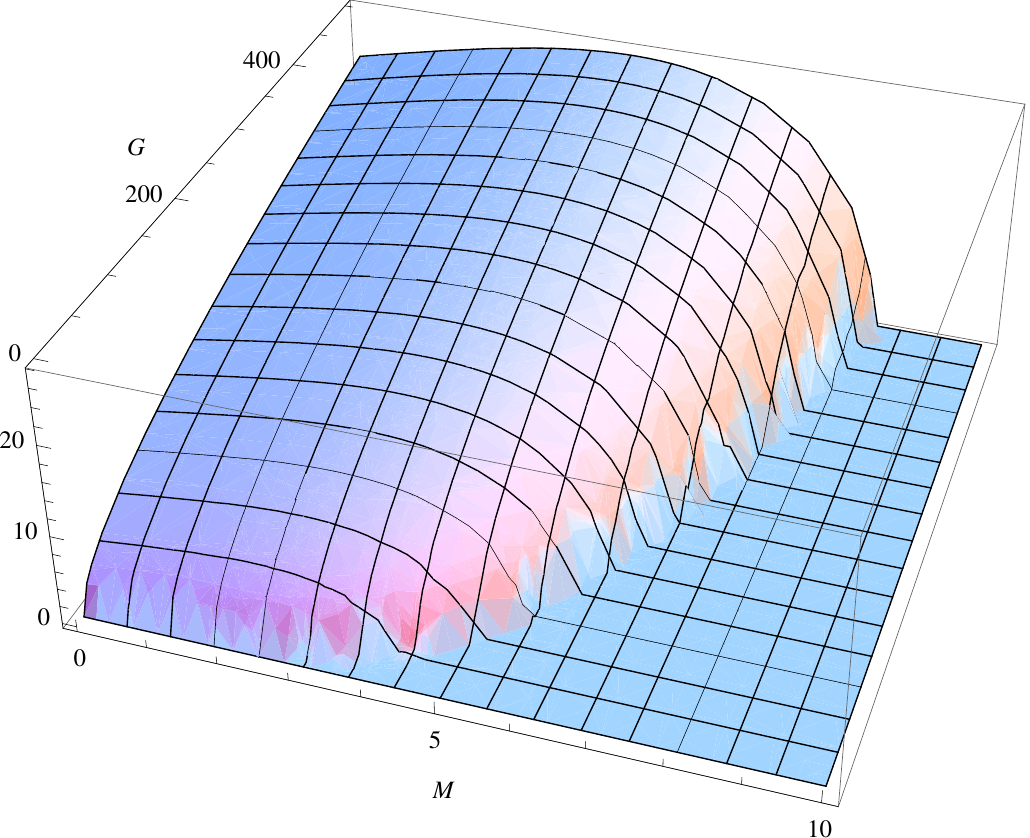}
\caption{Imaginary part of $m_2$ as a function of the CS mass $M$ and $G=\lambda g^8$ \label{imtwo}}
\end {figure}

\noindent Using computer algebra, it is straightforward to explicitly compute the four roots of \eqref{decomposition} and plot them in function of the parameters of the theory. Call $m_1$ and $m_2$ the masses that are real for $G < (3M/4)^4$, and $m_3$ and $m_4$ those that are always complex, and conjugate to each other, for any value of the parameters. The masses $m_3$ and $m_4$ never correspond to any physical excitation. The real parts of the masses $m_1$ and $m_2$ are plotted in \figurename s \ref{mone} and \ref{mtwo}. Both masses are positive, but are not real for all values of $M$ and $G$. Indeed, in order to explicitly show the threshold which divides parameter space in a region with and one without physical states, \figurename s \ref{imone} and \ref{imtwo} show the imaginary parts of the masses $m_1$ and $m_2$.

\begin{figure}
\centering
\includegraphics{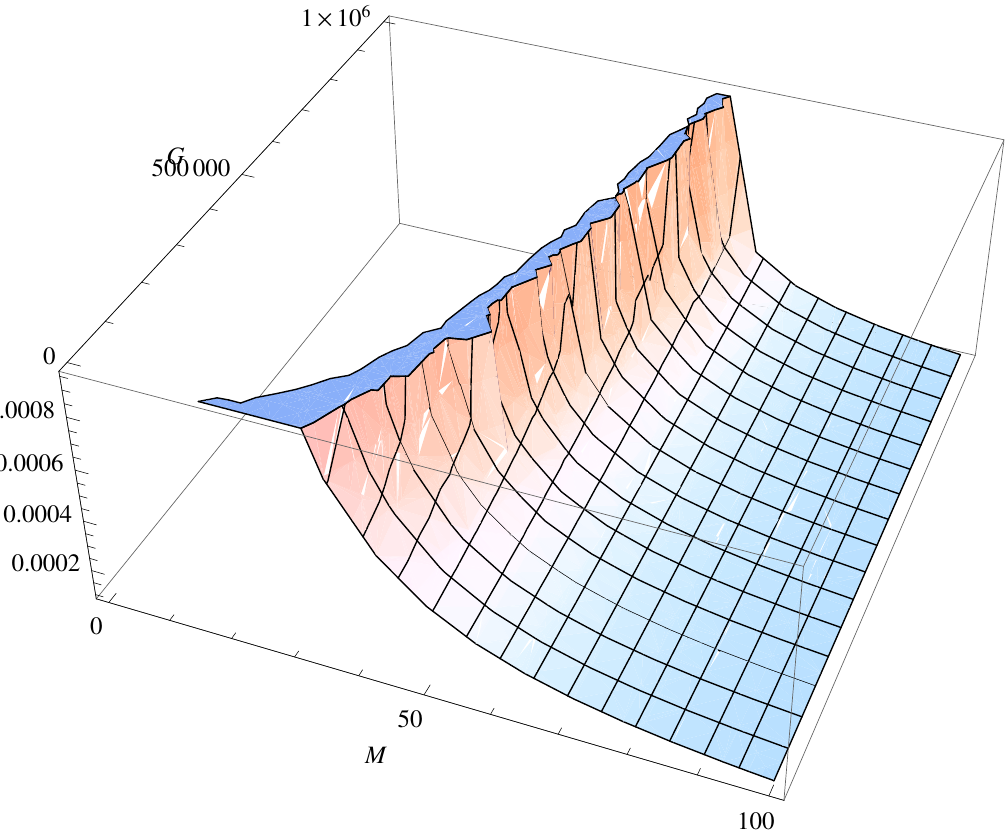}
\caption{$\mathcal{R}_{1}$ as a function of the CS mass $M$ and $G=\lambda g^8$ \label{resone}}
\label{Res1}
\end {figure}

\begin{figure}
\centering
\includegraphics{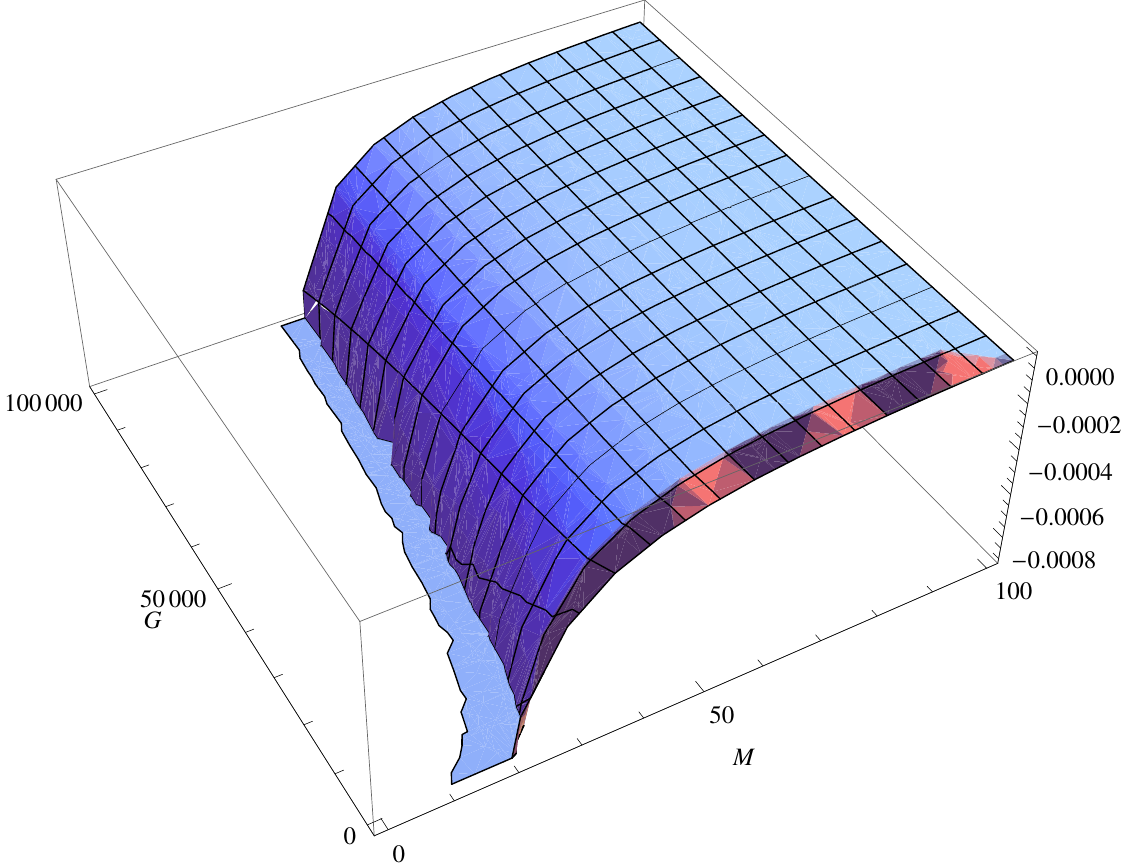}
\caption{$\mathcal{R}_{2}$ as a function of the CS mass $M$ and $G=\lambda g^8$\label{restwo}}
\label{Res2}
\end {figure}

\begin{figure}
\centering
\includegraphics{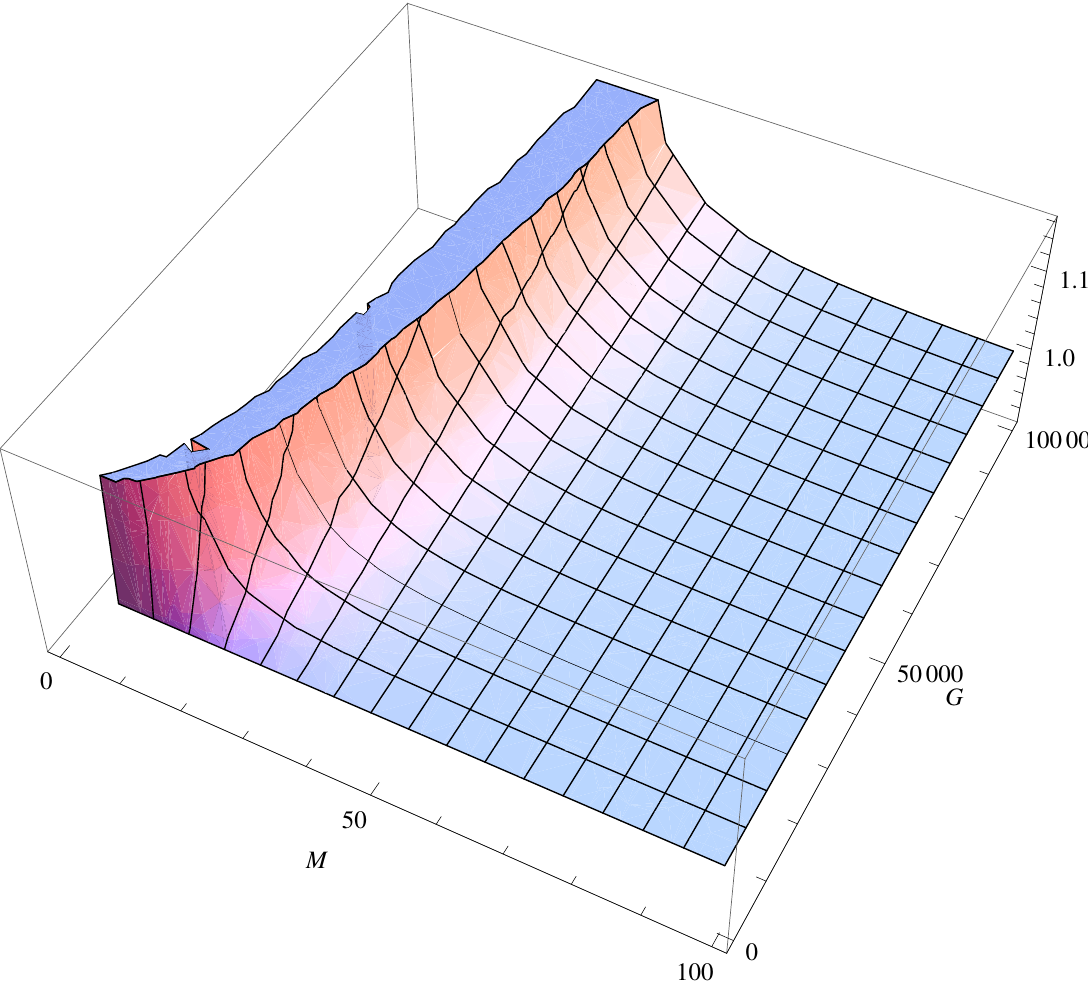}
\caption{$\mathcal{F}_{1}$ as a function of the CS mass $M$ and $G=\lambda g^8$\label{resparone}}
\end {figure}

\begin{figure}
\centering
\includegraphics{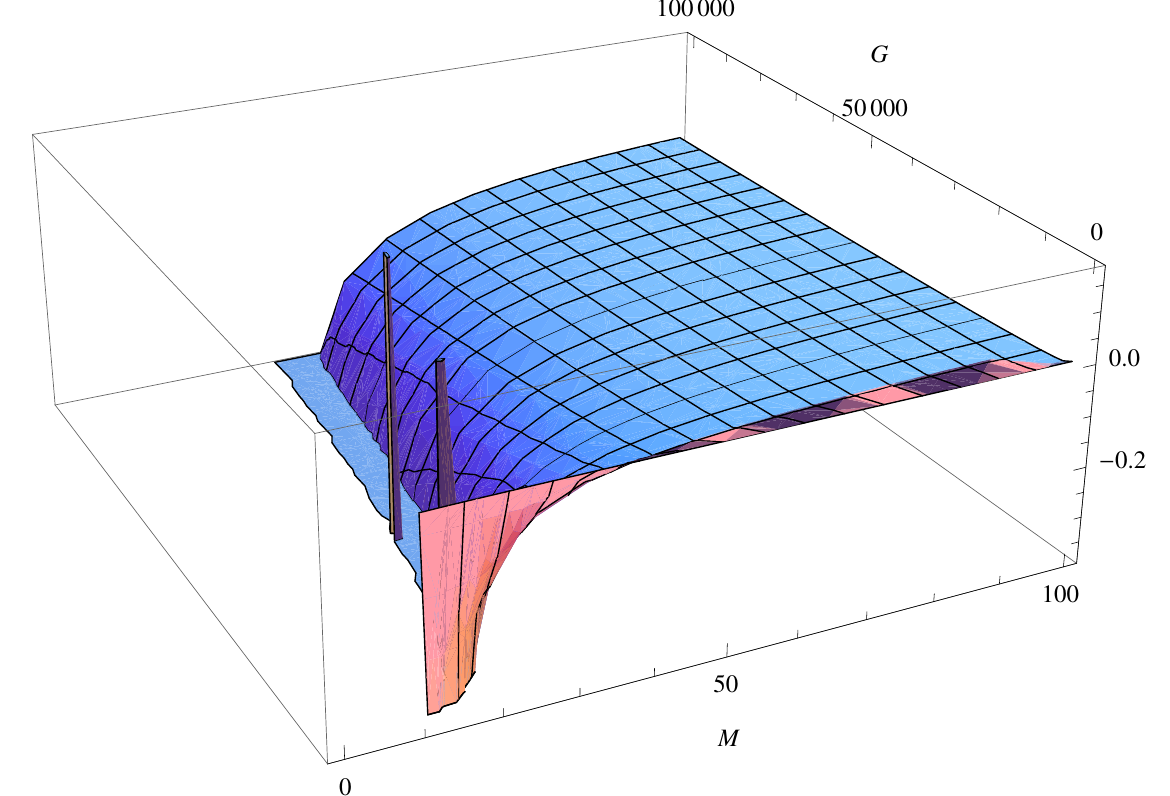}
\caption{$\mathcal{F}_{2}$ as a function of the CS mass $M$ and $G=\lambda g^8$\label{respartwo}}
\end {figure}

\noindent The masses of the parity violating part of the propagator are identical to the masses of the parity conserved part. The residues, though, are slightly different. \figurename s \ref{resone}, \ref{restwo}, \ref{resparone}, and \ref{respartwo} show their behavior in terms of the parameters $M$ and $G$. In both cases --- parity conserved and parity violating --- the residues corresponding to the massive pole $m_1$ --- $\mathcal{R}_{1}$ and ${\cal {F}}_{1}$ --- are positive, while the ones of $m_2$ --- $\mathcal{R}_{2}$ and $\mathcal{F}_{2}$ --- are negative. This means that the state associated with the mass $m_2$ cannot possibly represent a physical excitation.

\noindent Similar results were also found in \cite{Capri:2012cr} and \cite{Capri:2012ah}, where the Yang--Mills--Higgs theory was studied from the perspective of the Gribov problem. Despite the very different nature of the theory, a similar regime with two real masses, one with positive and one with negative residue, was also found there. Of course, we remind here that, when the weak coupling condition  \eqref{nogribov} is fulfilled, we have the standard particle spectrum, as given by the gauge propagator \eqref{PropCS}. Moreover, it has been shown in \cite{Canfora:2013kma} that the Gribov gap equation at finite temperatures in pure Yang--Mills theory produces a phase diagram which is very close to the ones obtained in  \cite{Capri:2012cr} and \cite{Capri:2012ah} in which the temperature plays the role of the Higgs vev.\\\\The present analysis might be relevant for the study of QCD at high temperatures. Indeed, as it is well known, in this case the theory can be described with an effective three-dimensional gauge theory in which the Chern--Simons term appear upon integrating out the fermions, see, for instance, \cite{Deser1} \cite{Deser2}, a detailed review being \cite{dunne}. The coupling constant of this kind of induced Chern--Simons term is proportional to the number of fermions flavours $N_f$. Hence, the present results imply that when the (a-dimensional combination of the Gribov parameter with the) Yang--Mills coupling is very small compared with the flavours number then the theory is not in the confining phase while when the (a-dimensional combination of the Gribov parameter with the) Yang--Mills coupling is very large compared with $N_f$ then the theory is in the confining phase.
These conclusions are very satisfactory from the intuitive point of view since it is well known that adding fermions flavours to Yang--Mills action ``decreases'' the confining character of the theory (see, for instance, \cite{veneziano}).

\section{Conclusion}

In this paper the Gribov semi-classical approach to eliminate gauge copies has been applied to Yang--Mills Chern--Simons theory in three dimensions. Unlike what happens in pure Yang--Mills theory, whose propagator is always confining at zero temperature within the Gribov semi-classical approach, the presence of the Chern--Simons topological term gives rise to a new regime in which a physical massive mode can propagate. In particular, the present analysis shows that there is a range of parameters, {\it i.e.} small Yang--Mills coupling constant and large values of the Chern--Simons coupling $M$, in which the theory is not in the confined phase since real poles corresponding to physical excitations appear. On the other hand, when the Yang--Mills coupling is large and the Chern--Simons coupling is small all the poles of the propagator are complex and the theory is in the confined regime. Therefore, even when the non-perturbative effects of the gauge copies are taken into account in the three-dimensional Yang--Mills--Chern--Simons theory, there is still a region of the parameters space corresponding to the Deser--Jackiw--Templeton massive gauge theory regime. Only when the Yang--Mills coupling is large enough compared to the Chern--Simons one, the confined phase appears.
The present analysis can be quite relevant in the study of QCD at high temperatures since, in this case, the theory can be described with an effective three-dimensional theory in which the Chern--Simons term appear upon integrating out the fermions. \\\\ Another issue worth to be investigated in the future is the possibility of implementing the restriction to the Gribov region to all orders, which would amount to construct a local Gribov-Zwanziger type action, as done in the case of pure 3d Yang-Mills see, for instance, ref.\cite{Dudal:2008rm}. In principle,  provided  the starting partition function is gauge-invariant, the terms which implement the restriction to the first Gribov region depend essentially only on the form of the gauge fixing itself. In this sense, one could implement the restriction to the first Gribov region beyond one-loop in a consistent way by adding to the starting action Zwanziger's horizon term in its local form \cite{Dudal:2008rm}. This would lead to a kind of local Gribov-Zwanziger action for Yang-Mills-Chern-Simons theory. Furthermore, it has been established that both Chern-Simons and Yang-Mills-Chern-Simons theories are ultraviolet finite \cite{Delduc:1990je,DelCima:1997pb}. It would be interesting to check if these finiteness properties would still hold in the presence of the horizon term. Finally, the formation of suitable lower dimensional dynamical condensates in a way similar to the so-called Refined-Gribov-Zwanziger action \cite{Dudal:2008rm} is also worth to be investigated.

\section{Acknowledgments}
This work of F. C. and A.G are partially supported by the FONDECYT grants N° 1120352 and N° 3130679. A. G. also wants to thanks the Max Planck Institute for his hospitality, where this manuscript was partially written. The Centro de Estudios Cientificos (CECs) is funded by the Chilean Government through the Centers of Excellence Base Financing Program of CONICYT.

\end{document}